\documentclass[aps,pra,preprint,groupedaddress,showpacs]{revtex4}
\usepackage{amsmath}
\usepackage{graphicx}
\begin{document}

\title{Time evolution of  decay of two identical quantum particles}

\author{Gast\'on Garc\'{\i}a-Calder\'on}
\email{gaston@fisica.unam.mx}
\author{Luis Guillermo Mendoza-Luna}
\affiliation{Instituto de F\'isica, Universidad Nacional Aut\'onoma de M\'exico,
Apartado Postal 20--364,  M\'{e}xico 01000, Distrito Federal, M\'exico }

\begin{abstract}
An analytical solution for the time evolution of decay of two identical non interacting quantum particles seated initially within a potential of finite range is derived using the formalism of resonant states.
It is shown that the wave function, and hence also the survival and nonescape probabilities, for
\textit{factorized symmetric}  and \textit{entangled  symmetric/antisymmetric} initial  states evolve in a distinctive
form along the exponentially decaying and nonexponential regimes. Our findings show the influence of the Pauli exclusion principle on decay.  We exemplify our results by solving exactly the  $s$-wave $\delta$ shell potential model.
\end{abstract}


\date{\today}

\maketitle

\section{Introduction}

Decay is one of the oldest topics in quantum theory. It  describes the escape
by tunneling of particles from a certain region.
In addition to the wave function itself, two quantities of interest are the survival and the nonescape probabilities.
As time evolves, the former is defined as the probability that a decaying particle remains in its original state
and the latter as the probability that the decaying particle remains within the interaction potential.

In this work, the wave function for the decay of  two identical non interacting particles is obtained both as a discrete expansion involving the full set of complex poles (and resonant states) of the outgoing Green's function to the problem and also as a discrete expansion that includes proper complex poles plus an integral contribution  which is more appropriate to study the long-time behavior. The above approaches have been used into the description of single-particle decay \cite{gc10,gcmv07}. There, in general, one finds, after an ultrashort non exponential contribution, an exponential decaying regime that depends on a single dominant resonance term followed at long times by an inverse power law as $t^{-3}$ ($s$ wave symmetrical potentials) \cite{gc10}. The short and  long-time nonexponential regimes have been experimentally verified in recent  years \cite{raizen97,rhm06}.

The escape behavior of identical particles is of interest, particularly since recent studies have shown that the Pauli exclusion principle has consequences upon the particular power law obeyed in the free time evolution (no potential)
of identical particles  confined initially in a finite region of space \cite{taniguchi11}.
The decay of a few-body Tonks-Girardeu gas has also been studied in this context by considering an integration over all the coordinates of the particles but one. There, it was also found an inverse-power law proportional to $t^{-3}$ \cite{dcdgcmr06}. The approach followed here considers both the survival and nonescape probabilities as  truly multiparticle observable quantities and as such the integrations will be all done in one single step.

The paper is organized as follows. Section \ref{formalism} reviews succinctly  the formalism of
resonant states,  in particular regarding the time evolution of single-particle decay. Section \ref{two particles}
provides a derivation of exact expressions for  the time evolution of decay of two identical particles as a resonance
expansion in terns of Moshinsky functions, both for  \textit{factorized symmetric} and \textit{symmetric/antisymmetric}
states and introduces the two-particle expressions for the survival and nonescape probabilities. Section \ref{ws}
discusses a model calculation involving the $s$ wave $\delta$ shell  potential which involves analytical expressions
for the two-particle solutions both along the exponential and  long-time nonexponential  regimes.
Finally, Section \ref{remarks} provides the concluding remarks.

\section{Formalism}\label{formalism}
Consider a single particle confined at $t=0$ along the internal region of a real spherically symmetrical finite-range potential, \textit{i.e.}, $V(r)=0$ for $r>a$. For simplicity we consider $s$ waves and choose as units
$\hbar=2m=1$. The solution to the time-dependent Schr\"odinger equation in the radial variable $r$, as an initial
value problem, may be written at time $t$ in terms of the retarded Green's function $g(r,r';t)$ of the problem as
\begin{equation}
\Psi(r,t)=\int_0^a {\! g(r,r^\prime,t)\Psi(r',0)\,\mathrm{d}r^\prime},
\label{eq:Psi_1_part}
\end{equation}
where $\Psi(r,0)$ stands for the arbitrary state initially confined within the internal interaction region.
In what follows, for the sake of the simplicity of the discussion, we refer to potentials that do not support
bound and antibound states.
A convenient form of the retarded time-dependent Green's function is expressed in terms of the
outgoing Green's function $G^+(r,r';k)$ of the problem.
Both quantities are related by a Laplace transformation
where the Bromwich contour  corresponds to a hyperbolic contour along the first quadrant of the $k$ plane \cite{gc10}
which may be evaluated by deforming the integration contour from $-\infty$ to $\infty$ along the
imaginary $k$ axis. This allows us to write the retarded Green's function as \cite{gc10}
\begin{equation}
g(r,r^\prime,t)=\sum_{p=-\infty}^{\infty}u_p(r)u_p(r')M(z_p),\qquad (r,r')^\dagger < a
\label{eq:g_exact}
\end{equation}
where the sum extends over the  complex poles of the problem,  the notation $(r,r')^\dagger$  means that
the point $r=r^\prime=a$ is excluded in the above expansion (otherwise it diverges) and the function $M(z_r)$, the
so called Moshinsky function, is defined as \cite{gc10}
\begin{equation}
M(z_r)=\frac{i}{2\pi}\int_{-\infty}^{\infty}\frac {{\rm e}^{-i\hbar k^2t/2m}}{k-\kappa_r}dk=
\frac{1}{2} w(iz_r),
\label{16c}
\end{equation}
where $z_r=-\exp(-i \pi /4)\kappa_rt^{1/2}$, with $r= {\pm p}$, and the function
$w(\zeta)=\exp(-\zeta^2)\rm{erfc(-i\zeta)}$ stands for the Faddeyeva or complex error function \cite{abramowitz} for which there exist efficient computational tools \cite{poppe90a}.
The functions $u_p(r)$ in (\ref{eq:g_exact}) correspond to the so called resonant  states
(also known as quasinormal modes) which are solutions to  the Schr\"{o}dinger equation of the problem obeying purely
outgoing boundary conditions which imply that the corresponding energy eigenvalues are complex, \textit{i.e.},
$E_p=\kappa_p^2= \mathcal{E}_p-\frac{i}{2}\Gamma_p$, where $\mathcal{E}_p$ yields the
resonance energy of the decaying fragment and $\Gamma_p$ stands for the resonance width,  which yields the lifetime
$\tau_p=1/\Gamma_ p$ of a given resonance level. The lifetime of the system is defined by the longest lifetime,
\textit{i.e.}, the shortest width.
The complex poles $\kappa_p=a_p-ib_p$ are distributed along the third and fourth quadrants of the complex
$k$ plane in a well known manner \cite{newton}.

For proper poles, \textit{i.e.}, $a_p > b_p$, Eq. (\ref{eq:g_exact}) may be written alternatively, using some properties of the Faddeyeva function, as \cite{gc10}
\begin{equation}
g(r,r^\prime,t)= \sum_{p=1}^\infty u_p(r)u_p(r^\prime)e^{-i\mathcal{E}_pt}e^{-\frac{1}{2}\Gamma_pt}+ I(r,r';t)
\label{eq:exp_prop}
\end{equation}
where
\begin{equation}
I(r,r';t)=\\ \sum_{p=1}^{\infty} [u_p(r)u_p(r')M(-z_p) - u^*_p(r)u^*_p(r')M(z_{-p}],
\label{ine}
\end{equation}
and we recall that $(r,r')^\dagger < a$ .
Since the potential is real, it follows  from time-reversal invariance that
$u_{-p}(r)=u_p^*(r)$ and $\kappa_{-p}=-\kappa_p^*$.
The last term in (\ref{eq:exp_prop}) becomes relevant both at ultrashort and asymptotic long times
\cite{gc10}. The description  of ultrashort times is more involved and will not be considered here since
it possibly requires different considerations.

There is another route to analyze the long-time behavior of $g(r,r';t)$ which follows by closing the Bromwich contour
mentioned above along a straight line $C_l$ that is $45^{\circ}$ off the real axis and goes through the
origin \cite{gc10}. The resulting expression, that is equivalent to Eq. (\ref{ine}), reads
\begin{equation}
I(r,r';t)=(i/\pi)\int_{C_l} G^+(r,r';k)\exp(-ik^2t)2kdk.
\label{insd}
\end{equation}
It turns out that this integral term may be evaluated at long times by the steepest descent method as it has a saddle point at $k=0$ and hence one may perform a Taylor expansion of $G^+(r,r^\prime,k)$ around that value to evaluate the integral \cite{gc10}. Thus, alternatively, at long times, the retarded propagator may also be written as
\begin{equation}
g(r,r^\prime,t)\approx\sum_{p=1}^\infty u_p(r)u_p(r^\prime)e^{-i\mathcal{E}_pt}e^{-\frac{1}{2}\Gamma_pt}+
\\ \sum_{m=1}^\infty \frac{\eta_m}{t^{(2m+1)/2}}\frac{\partial^{2m-1}}{\partial k^{2m-1}}G^+(r,r^\prime,k)\Big{|}_{k=0}
\label{eq:desarr_prop_g_asint}
\end{equation}
where $\eta_1=1/\sqrt{4\pi i}$, $\eta_2 = -\sqrt{i/(64\pi)}$ and $\eta_3 = -1/\sqrt{4096\pi i}$.
For decay of a single particle it suffices to take $m=1$ in Eq. (\ref{eq:desarr_prop_g_asint}). However, already for
two particles, higher values of $m$ are required, as discussed below.

\section{Two identical particles}\label{two particles}

In the case of a system of identical  non interacting particles, it is known that the Hamiltonian $H$ must be symmetric under the permutation of the indices of the particles so the exchange operator and $H$ necessarily commute. Thus, it is enough to impose the appropriate symmetry/antisymmetry on the initial state $\Psi(y_1,y_2,0)$ since symmetry is conserved as time evolves. Hence, the time evolution for decay of two identical particles may be written as
\begin{equation}
\Psi(\mathbf{r},t)= \int_0^a {\!\int_0^a {\!g(r_1,y_1,t)g(r_2,y_2,t)\Psi(\mathbf{y},0)\,\mathrm{d}y_1}\,\mathrm{d}y_2},
\label{eq:Psi_2_parts}
\end{equation}
where $\mathbf{r}$ and $\mathbf{y}$ denote, respectively,  $(r_1,r_2)$ and  $(y_1,y_2)$.

A simple choice, which corresponds to a symmetric state, is given by the product
of single particle states $\psi_\alpha(y_1)$ and $\psi_\alpha(y_2)$, with $\alpha$ denoting the state,
\begin{equation}
\Psi(\mathbf{y},0)=\psi_\alpha(y_1)\psi_\alpha(y_2).
\label{fac_sim}
\end{equation}
Substitution of (\ref{fac_sim}) and (\ref{eq:g_exact}) into
(\ref{eq:Psi_2_parts}) yields the \textit{factorized symmetric} state
\begin{eqnarray}
\Psi(\mathbf{r},t)&=&  \sum_{p,q=-\infty}^\infty C_{p,\alpha}C_{q,\alpha}
u_p(r_1)u_q(r_2)M(z_p)M(z_q)=\nonumber \\[.3cm]
&&\left(\sum_{p=-\infty}^\infty C_{p,\alpha}u_p(r_1)M(z_p)\right)
\left(\sum_{q=-\infty}^\infty C_{q,\alpha}u_q(r_2)M(z_q)\right)
\label{fac_Psi_exact}
\end{eqnarray}
where $(r_1, r_2) \leq a$ and $C_{n,\alpha}$, with $n=p,q$, is given by
\begin{equation}
C_{n,\alpha}=\int_0^a {\!u_n(y)\psi_\alpha(y)\,\mathrm{d}y}.
\label{coefs}
\end{equation}
Another choice for the initial  state consists of the linear combination of single-particle states $\psi_s(y_1)$
and $\psi_s(y_2)$.  Here, $s=\alpha,\beta$ refers to the possible states of the two particles, and hence we may write
\begin{equation}
\Psi(\mathbf{y},0)=\frac{1}{\sqrt{2}}(\psi_\alpha(y_1)\psi_\beta(y_2)\pm \psi_\beta(y_1)\psi_\alpha(y_2)),
\label{eq:e_inicial}
\end{equation}
where respectively,  the plus sign refers to \textit{entangled symmetric} and the minus sign to
\textit{entangled antisymmetric} states.
Then, substitution of (\ref{eq:e_inicial}) and (\ref{eq:g_exact}) into (\ref{eq:Psi_2_parts}) yields
\begin{equation}
\Psi(\mathbf{r},t)=  \frac{1}{\sqrt{2}}\sum_{p,q=-\infty}^\infty (C_{p,\alpha}C_{q,\beta} \pm C_{p,\beta}C_{q,\alpha}) u_p(r_1)u_q(r_2)M(z_p)M(z_q),
\label{eq:desarr_Psi_exact}
\end{equation}
where $(r_1,r_2) \leq a$ and the coefficients $C_{n,\beta}$  follow by replacing $\alpha$ for $\beta$ in (\ref{coefs}).
It is worth recalling that the coefficients $\{C_{n,s}\}$, which involve only single-particle states,
fulfill the relationship \cite{gc10}
\begin{equation}
{\rm Re} \left( \sum_{n=1}^\infty C_{n,s}{\bar C}_{n,s} \right )=1,
\label{sumrule}
\end{equation}
where ${\bar C}_{n,s}$ is defined as (\ref{coefs}) with $\psi_s(y)$ substituted by $\psi^*_s(y)$. Hence for real
initial states,  ${\bar C}_{n,s}=C_{n,s}$. Although the $C_{n,s}$ are complex and its real part may be negative,
they play a most relevant role in time-dependent expansions as discussed in Refs. \cite{gcrv09,gc10}.

Alternatively, one may consider  Eq. (\ref{eq:desarr_prop_g_asint}) instead of (\ref{eq:g_exact}) to calculate
Eqs. (\ref{fac_Psi_exact}) and (\ref{eq:desarr_Psi_exact}). This last procedure
provides explicit analytical expressions for the exponential decaying and long time inverse power terms.
We shall consider both possibilities for the model calculation below.

\subsection{Survival and nonescape probabilities}\label{probs}

The survival amplitude of a two-particle system is defined as
\begin{equation}
\label{eq:def_amplit_A}
A(t)=\int_0^a {\! \int_0^a {\! \Psi^*(r_1,r_2,0)\Psi(r_1,r_2,t)\,\mathrm{d}r_1}\,\mathrm{d}r_2};
\end{equation}
hence, the survival probability is given by
\begin{equation}
\label{eq:def_prob_S}
S(t)=|A(t)|^2.
\end{equation}
The nonescape probability of a two-particle system is defined as
\begin{equation}
\label{eq:def_prob_P}
P(t)=\int_0^a {\! \int_0^a {\! |\Psi(\mathbf{r},t)|^2\,\mathrm{d}r_1}\,\mathrm{d}r_2}.
\end{equation}
Once $\Psi(\mathbf{r},t)$ is known, the calculation of $S(t)$ and $P(t)$ follows from  Eqs. \eqref{eq:def_prob_S} and \eqref{eq:def_prob_P}.

Notice that $P(t)$ is always larger than $S(t)$. This is a general feature for these quantities,
that holds also in single-particle decay, which follows from their definition using the Cauchy-Schwarz 
inequality, \textit{i.e.}, in general $P(t) \geq S(t)$.

\section{Model}\label{ws}

The $\delta$ shell potential, whose mathematical simplicity allows it to describe the essential
physical features of the time evolution of decay \cite{gc10,winter61}, is suitable for calculations and extends
the work done for the free-particle case. For two particles this potential may be written as
\begin{equation}
\label{eq:def_poten_V}
V(r_1,r_2)=\lambda\delta(r_1-a)+\lambda\delta(r_2-a).
\end{equation}
For the single-particle case the resonant states of the corresponding problem satisfy the Schr\"{o}dinger equation with complex energy eigenvalues. They read
\begin{eqnarray}
u_p(r) \: &=& \: \left \{
\begin{array}{cc}
A_p\sin(\kappa_p r) & ;\,\, r \leq a \\[.3cm]
B_p\exp(i\kappa_pr) & ;\,\, r \geq a
\end{array}
\right.
\label{rs}
\end{eqnarray}
and  are normalized according to the condition
\begin{equation}
\int_0^au_p^2(r)dr + i u_p^2(a)/2\kappa_p=1.
\label{nc}
\end{equation}

From the usual boundary conditions for a $\delta$ potential the complex eigenvalues satisfy \cite{gc10}
\begin{equation}
2i\kappa_p+\lambda(e^{2i\kappa_pa}-1)=0.
\label{poles}
\end{equation}
There are well established procedures to calculate the complex poles $\kappa_n$ solving (\ref{poles}) \cite{gcmv07,gc10}.

A convenient feature of this model is that the outgoing Green function $G^+(r,r^\prime,k)$ may be written as the simple analytical expression
\cite{gcmv07}
\begin{equation}
G^+(r,r\,';k)=  -\frac{\sin(kr)}{k}
\left [\frac{\exp(ikr{\,'})-(\lambda/k) \sin (\,k(r{\,'}-a)\,)\exp(ika) }{1+(\lambda/k) \sin(ka) \exp(ika)} \right ],
\label{exactg}
\end{equation}
and hence, the partial derivatives of $G^+(r,r';k)$ that appear in Eq. \eqref{eq:desarr_prop_g_asint} may also be written explicitly. They are given in the Appendix \ref{app} and are used to obtain the expressions for the time-dependent wave solutions discussed below.
\begin{figure}[htb!]
\centering
\includegraphics[width=9cm]{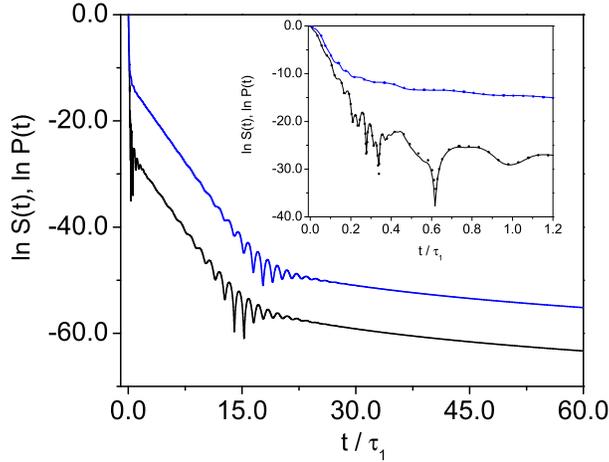} \caption{(Color online)  Plot of the ${\rm ln}\, S(t)$ (lower solid line)
and ${\rm ln}\, P(t)$ (upper solid line) as a function of time in lifetime units for the \textit{factorized symmetric
state} using  Eq. (\ref{fac_Psi_exact}). The inset shows a calculation of the same quantities at short times
using the purely exponential contributions (solid lines) and its corresponding comparison with exact calculations
in terms of Moshinsky functions (dotted lines). See text.}
\label{fig1}
\end{figure}

As initial states we choose  appropriate combinations of infinite box states
\begin{equation}
\psi_s(y) = \sqrt{\frac{2}{a}}\sin \left (\frac{s\pi y}{a} \right),
\label{is}
\end{equation}
with $s=(\alpha, \beta)$. Specifically, for the \textit{factorized symmetric} state (\ref{fac_sim}) we choose
the  product of two infinite box  states with $\alpha=6$ whereas for the \textit{ entangled symmetric/antisymmetric states} (\ref{eq:e_inicial}) we take $\alpha=1$ and $\beta=6$.

It is worth mentioning that the state of the system in the Hilbert space corresponds to a ket that can be factorized
into a space and a spin part as we are assuming a spin-independent Hamiltonian. The spin part of the state ket has
to be taken into account when considering the total symmetry of the state. In this work we will address only the spatial part of the state ket. We shall consider calculations involving both the exact analytical and asymptotic
expressions for the retarded propagator.

The calculations below are made using both the exact expressions (\ref{fac_Psi_exact}) and (\ref{eq:desarr_Psi_exact})
and the corresponding approximate expressions that exhibit the long time behavior explicitly. In the next subsection
we provide explicit analytical expressions for the corresponding  symmetric and antisymmetric wave functions.

The potential parameters employed in the calculations are:  $\lambda=6$ and $a=1$ and it was sufficient to consider $20$ poles.   

\subsection{Symmetric space wave functions}\label{swf}

We first consider the \textit{factorized symmetric}  wave function given by Eq. (\ref{fac_Psi_exact}).
It follows that the long time contribution of  $g(r,r^\prime,t)$, given by the second term on the right-hand side of
(\ref{eq:desarr_prop_g_asint}), may be truncated at order $t^{-3/2}$ and thus the wave function reads

\begin{eqnarray}
\Psi(\mathbf{r},t)&\approx& \sum_{p,q=1}^\infty C_{p,\alpha}C_{q,\alpha}u_p(r_1)u_q(r_2)e^{-i(\mathcal{E}_p+\mathcal{E}_q)t} e^{-\frac{1}{2}(\Gamma_p+\Gamma_q)t}-\frac{r_1r_2D^2_\alpha \eta_1^2}{(1+\lambda a)^4t^3}-\nonumber \\ [.3cm]
&&\frac{i\eta_1}{(1+\lambda a)^2t^{3/2}} D_\alpha \sum_{p=1}^\infty { C_{p,\alpha}(r_2 u_p(r_1)+r_1 u_p(r_2))e^{-i\mathcal{E}_pt}e^{-\frac{1}{2}\Gamma_p t} },
\label{exp_fac_Psi_sim}
\end{eqnarray}
where $(r_1,r_2) \leq a$ and we have used the shorthand $D_{\alpha}= \int_0^a {\!y\,\psi_\alpha(y)\,\mathrm{d}y}$.
Equation \eqref{exp_fac_Psi_sim} is a description up to leading terms of the solution; it has essentially three contributions: purely exponential decaying terms, a purely non-exponential inverse power law and a mixed term, made up by several inverse-power multiplied by exponential contributions.
Figure \ref{fig1} exhibits a plot of both ${\rm ln}\, S(t)$ and ${\rm ln}\,P(t)$ in units of the lifetime of the system, (which is given by $\tau_1=1/\Gamma_1$, since $\Gamma_1$ is the shortest decay width).
We observe several regions of interest for both the survival and nonescape probabilities: for short times there is an exponential regime in which the slope of ${\rm ln}\,S(t)$  and ${\rm ln}\,P(t)$ goes as  $-2\Gamma_6$,
then, there is a change of the slopes into $-2\Gamma_1$ (since this is the exponential term with the longest duration);
this is followed by interference contributions between the purely exponential terms and the purely
inverse power term and finally the non-exponential asymptotic regime that goes as $t^{-6}$.

\begin{figure}[htb!]
\centering
\includegraphics[width=9cm]{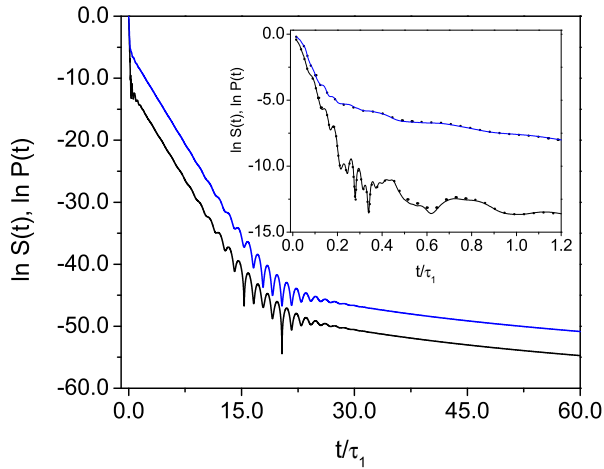}
\caption{(Color online) Plot of the ${\rm ln}\, S(t)$ (lower solid line) and ${\rm ln}\, P(t)$ (upper solid line)
as a function of time in lifetime units for the \textit{entangled symmetric state} using
Eq. (\ref{eq:desarr_Psi_sim}). The inset shows a calculation at short times of the same quantities using the
purely exponential contributions (solid lines) and the corresponding comparison with exact calculations in terms
of Moshinsky functions (dotted lines).
See text.}
\label{fig2}
\end{figure}

We now consider the \textit{entangled symmetric} wave function (plus sign) given by Eq. (\ref{eq:e_inicial}).
It follows that  the long time contribution of $g(r,r^\prime,t)$, given by the second term on the right-hand side of
(\ref{eq:desarr_prop_g_asint}), may be truncated at order $t^{-3/2}$ and thus the wave function reads
\begin{widetext}
\begin{multline}
\Psi(\mathbf{r},t)\approx\frac{1}{\sqrt{2}}\sum_{p,q=1}^\infty (C_{p,\alpha}C_{q,\beta}+C_{p,\beta}C_{q,\alpha})u_p(r_1)u_q(r_2)e^{-i(\mathcal{E}_p+\mathcal{E}_q)t} e^{-\frac{1}{2}(\Gamma_p+\Gamma_q)t}-\frac{\sqrt{2}r_1r_2D_\alpha D_\beta\eta_1^2}{(1+\lambda a)^4t^3}\\ -\frac{i\eta_1}{\sqrt{2}(1+\lambda a)^2t^{3/2}} \sum_{p=1}^\infty { (D_\beta C_{p,\alpha}+D_\alpha C_{p,\beta})(r_2 u_p(r_1)+r_1 u_p(r_2))e^{-i\mathcal{E}_pt}e^{-\frac{1}{2}\Gamma_p t} },
\label{eq:desarr_Psi_sim}
\end{multline}
\end{widetext}
where we recall that $(r_1,r_2) \leq a$.
We note that Eq. \eqref{eq:desarr_Psi_sim} is again a description up to leading terms of the solution; it has essentially three contributions: purely exponential decaying terms, a purely non-exponential inverse power law and a mixed term, made up by several inverse-power multiplied by exponential contributions. In Fig. \ref{fig2} we  see several regions of interest for both ${\rm ln}\,S(t)$ and ${\rm ln}\,P(t)$: for short times there is an exponential regime in which the slope of ${\rm ln}\,S(t)$  and ${\rm ln}\,P(t)$ goes as  $-(\Gamma_1+\Gamma_6) \approx -\Gamma_6$ (since $\Gamma_6 \gg \Gamma_1$), then, there is an interference among the purely exponential terms and between them and the mixed term; the next region goes with slope $-2\Gamma_1$; then there is an interference between the purely exponential terms and the purely inverse power term and finally the non-exponential asymptotic regime that goes as $t^{-6}$.
\subsection{Antisymmetric space wave function}\label{awf}

The \textit{entangled antisymmetric} wave function follows by choosing
the minus sign in (\ref{eq:e_inicial}) and proceeding in a similar fashion as above. At long times, the propagator $g(r,r^\prime,t)$ has to be expanded up to $t^{-7/2}$ order since the terms of order $t^{-3}$ and $t^{-4}$ in $\Psi(\mathbf{r},t)$ cancel out exactly. Thus, we obtain a time-development given by
\begin{widetext}
\begin{multline}
\Psi(\mathbf{r},t)\approx\frac{1}{\sqrt{2}}\sum_{p,q=1}^\infty (C_{p,\alpha}C_{q,\beta}-C_{p,\beta}C_{q,\alpha})u_p(r_1)u_q(r_2)e^{-i(\mathcal{E}_p+\mathcal{E}_q)t} e^{-\frac{1}{2}(\Gamma_p+\Gamma_q)t}+ \\\left(\eta_2^2-\frac{10\eta_1\eta_3}{3}\right)
\frac{(r_1^3r_2-r_2^3r_1)(D_\beta G_\alpha-G_\beta D_\alpha)}{\sqrt{2}(1+\lambda a)^4t^5} - \\
\frac{i\eta_1 }{\sqrt{2}(1+\lambda a)^2t^{3/2}}\sum_{p=1}^\infty (D_\beta C_{p,\alpha}-D_\alpha C_{p,\beta})(r_2u_p(r_1)-r_1u_p(r_2))e^{-i\mathcal{E}_pt}e^{-\frac{1}{2}\Gamma_p t}
\label{eq:desarr_Psi_antisim}
\end{multline}
\end{widetext}
where $(r_1,r_2) \leq a$ and $G_{\alpha} = \int_0^a {\!y^3\,\psi_\alpha(y)\,\mathrm{d}y}$.
\begin{figure}[htb!]
\centering
\includegraphics[width=9cm]{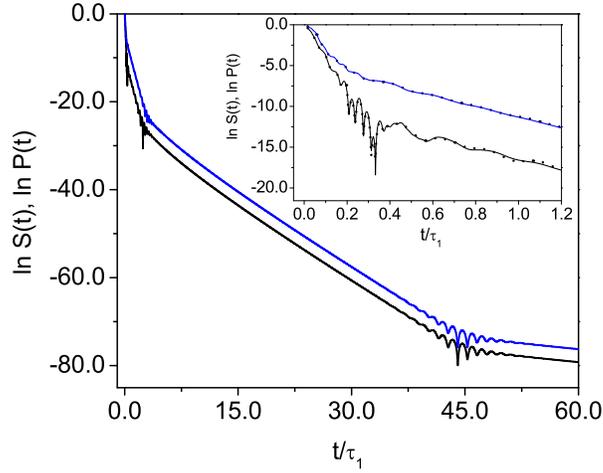}
\caption{(Color online) Plot of the ${\rm ln}\,S(t)$ (lower solid line) and ${\rm ln}\, P(t)$ (upper solid line) as a
function of time in lifetime units for the \textit{entangled antisymmetric state} using  Eq. (\ref{eq:desarr_Psi_antisim}).
The inset shows a calculation of the same quantities at short times using the purely exponential contributions
(solid lines) and the corresponding comparison with exact calculations in terms of Moshinsky functions (dotted lines).
See text.}
\label{fig3}
\end{figure}
Note that Eq. \eqref{eq:desarr_Psi_antisim} has also three leading contributions: a purely exponential decaying one, a purely non-exponential inverse power law (now of order $t^{-5}$) and again a mixed term. It is relevant to realize that $(C_{p,\alpha}C_{q,\beta}-C_{p,\beta}C_{q,\alpha})$ enforces an exact cancelation of the terms in the sum whenever $p=q$ and hence there are not contributions of order $-2\Gamma_p$, for any $p$,  in the exponential
decaying terms as occurs in  the symmetric case.
In Fig. \ref{fig3} we can identify several characteristic regions for ${\rm ln}\,S(t)$ and ${\rm ln}\,P(t)$:
for short times there is an exponential regime in which the slope of both ${\rm ln}\,S(t)$ and ${\rm ln}\,P(t)$ goes as
$-(\Gamma_1+\Gamma_6) \approx -\Gamma_6$ (recalling that $\Gamma_6 \gg \Gamma_1$); next one observes a short linear region of slope
$-(\Gamma_1+\Gamma_2)$; then it follows an interference term between the purely exponential and  the mixed
contributions; the next region comes from the mixed terms and goes with slope $-\Gamma_1$; then there is an
interference contribution between  the mixed terms and the inverse power term, and finally the  non-exponential
regime as $t^{-10}$.

A salient feature of the insets in Figs. \ref{fig1}, \ref{fig2} and \ref{fig3} is that the survival
probabilities are much less than the corresponding nonescape probabilities. This also occurs for single-particle 
decay \cite{gcmv07}. It occurs whenever the decay process starts from an excited state. Otherwise along the 
exponentially decaying regime $S(t)$ and $P(t)$  are practically indistinguishable. 

In the absence of a potential the problem reduces to the free time evolution of an initially confined
two particle state. This has been recently discussed in Ref. \cite{taniguchi11}. Their results are reproduced
by letting $\lambda=0$ in  Eqs.(\ref{eq:desarr_Psi_sim}) and  (\ref{eq:desarr_Psi_antisim}), in which case there
are no pole contributions and the symmetrical and antisymmetrical free  wave solutions evolve respectively as
$t^{-3}$ and $t^{-5}$.

\section{Concluding remarks}\label{remarks}

This work shows that the character of the time evolution of decay of two identical non-interacting particles depends on whether the initial state is \textit{factorized symmetric} or \textit{entangled symmetric/antisymmetric}. The results obtained indicate that each of the above initial cases exhibits in general a distinctive behavior along the exponential and long-time nonexponential regimes. The differences in the exponential decaying regime may be observed in the different slopes of the plots discussed here and, for the asymptotic long-time regime,  in the different inverse power laws obeyed.

We have restricted the discussion to expansions of the time-dependent solution along the internal interaction region. This is sufficient to calculate the survival and nonescape probabilities. One might also consider the time evolution for decay along the external interaction region by generalizing to two particles the resonant expansion discussed  in Ref. \cite{gc10}, which yields for the single-particle time evolving retarded Green's function the expression
\begin{displaymath}
g(r,r^\prime,t)=\sum_{p=-\infty}^{\infty}u_p(a)u_p(r')M(z_p), \quad r' < a, \quad r \geq a,
\end{displaymath}
where the argument $z_p$ of the Moshinsky function now reads $z_p=\exp(-i\pi/4)(1/2t^{1/2}[((r-a)-2\kappa_pt]$.
On the other hand, one may extend the discussion given here to include the effect of bound states just by adding the
corresponding bound and antibound  terms to the  resonance sums (\ref{fac_Psi_exact}) and
(\ref{eq:desarr_Psi_exact}) \cite{gc10}. This may lead to interesting effects as `trapping' \cite{dmagc05}.

As our example exhibits,  the survival and nonescape probabilities both go at asymptotically long times, respectively,  as $t^{-6}$ for both \textit{factorized/entangled symmetric} states, and as $t^{-10}$ for
\textit{entangled antisymmetric} states. In the limit of a vanishing interaction potential our results
tend to  the free evolving case discussed in Ref. \cite{taniguchi11}. The above long-time results are
in contrast with the $t^{-3}$ behavior for decay of a single particle \cite{gc10}.
Evidently, the exponential decaying regime is the one more accessible to experimental verification. Here the \textit{factorized symmetric} states possess a different behavior than for the \textit{entangled symmetric/antisymmetric} states, which, on the other hand, exhibit at short times a similar decay rate.

We hope that the distinct behaviors for the time evolution of decay discussed here
might be  experimentally verified in quantum systems where one may manipulate to certain degree the potential parameters such as in cold atoms \cite{raizen97}.

\textit{Note-} After submission of this work, we became aware of Ref. \cite{delcampo11} which develops an approach to many particle decay addressing only the long-time decay regime. 

\begin{acknowledgments}
G.G.-C. acknowledges A. del Campo for calling our attention to the work of Ref. \cite{taniguchi11}
and the partial financial support from DGAPA-UNAM under Grant IN 112410. Both authors acknowledge 
useful discussions with S. Cordero.
\end{acknowledgments}
\appendix
\section{Derivatives of $G^+(r,r';k)$ with respect to $k$ evaluated at $k=0$} \label{app}

We write down below the explicit expressions for the derivatives with respect to $k$ of $G^+(r,r';k)$  at $k=0$ which
are required to obtain the analytical expressions for the time-dependent solutions given in Subsecs. \ref{swf} and \ref{awf}. Here we  recall that  $(r,r')^\dag < a$.

\begin{eqnarray}
\frac{\partial}{\partial k}G^+(r,r^\prime,k)\Big{|}_{k=0} &=&  -\frac{irr^\prime}{h_1}  \label{eq:1a} \nonumber \\[0.3cm]
\frac{\partial^3}{\partial k^3}G^+(r,r^\prime,k)\Big{|}_{k=0} &=&  \frac{irr^\prime(h_1(r^2+(r^\prime)^2)-h_2)}{h_1^2} \nonumber
\label{eq:3a}
\end{eqnarray}
\begin{widetext}
\begin{equation}
\frac{\partial^5}{\partial k^5}G^+(r,r^\prime,k)\Big{|}_{k=0} =  -\frac{irr^\prime }{3h_1^3}\left(h_1^2(3r^4+3(r^\prime)^4+10r^2(r^\prime)^2)+ h_3(r^2+(r^\prime)^2)+h_4\right) \nonumber
\label{eq:5a}
\end{equation}
\end{widetext}
where

\begin{eqnarray}
h_1&=&(1+\lambda a)^2 \label{eq:def_coef_h_1} \nonumber \\[0.3cm]
h_2&=& 8\lambda a^3+2\lambda^2a^4 \label{eq:def_coef_h_3}\nonumber\\[0.3cm]
h_3&=& -120\lambda^3 a^5-80\lambda a^3-20\lambda^4 a^6-180\lambda^2a^4 \label{eq:def_coef_h_4}\nonumber\\[0.3cm]
h_4&=& 192\lambda^3 a^7-96\lambda a^5+24\lambda^4 a^8+432\lambda^2 a^6 \label{eq:def_coef_h_5} \nonumber
\end{eqnarray}
%

%
%\bibliography{references}
%\bibliography{refs}

\end{document}